\documentclass{ckm}                 

\usepackage{txfonts}            

\confname{Workshop on the CKM Unitarity Triangle, IPPP Durham, April
  2003}

\title{Lattice Determination of Semileptonic Form Factors}

\author{T. Onogi}

\address{Yukawa Institute for Theoretical Physics, Kyoto University,
	 Kyoto, 606-8502, Japan}

\newcommand{\boldvec}[1]{\mbox{\boldmath$#1$}}

\newcommand{\Btopi}{B\rightarrow\pi l\nu}
\newcommand{\Dtopi}{D\rightarrow\pi l\nu}
\newcommand{\Btorho}{B\rightarrow\rho l\nu}
\newcommand{\BtoD}{B\rightarrow D^{(*)} l\nu}
\newcommand{\BtoK}{B\rightarrow K^{(*)} l^+ l^-}
\newcommand{\vdotk}{v \cdot k_{\pi}}

\begin{document}

\begin{abstract}We report on the Lattice determination of the
semileptonic form factors by lattice QCD. Comparison with the 
light-cone QCD sum rules are made for $\Btopi, \Btorho$ semileptonic 
decays.
\end{abstract}

\maketitle


\section{Introduction}

Computation of the semileptonic form factors 
for $B\rightarrow\pi(\rho )l\nu$, $\BtoD$ and $\BtoK$ decays is 
a key to the determination of the CKM matrix elements 
$|V_{\rm ub}|$, $|V_{\rm cb}|$ and $|V_{\rm ts}|$ .
In this report, I review the recent results by 
lattice QCD methods.  
I will mainly focus on the semileptonic decay $\Btopi$, 
because lattice calculations in various approaches are already
at hand~\cite{Bpi_APE,Bpi_UKQCD,Bpi_FNAL,Bpi_JLQCD,Bpi_NRQCD} 
and because most of the problems which exists in this decay
are common to other processes. 

Lattice QCD calculations of $\Btopi$ form factors suffer from 
(1) the error from the large heavy quark mass $m_Q$ in lattice unit, 
(2) the chiral extrapolations in the light quark mass $m_q$ 
    using the simulation results with relatively large mass region 
    $m_q> m_s/2$, 
and (3) the limitations from the accessible kinematic range 
of $q^2$ from statistical and systematic errors.

In order to solve the first problem, two different approaches are made.
One is to avoid the large discretization errors of $O(a m_Q)$
by computing the form factors with the conventional relativistic quark 
action for charm quark mass region, and then extrapolate the results 
in $1/m_Q$. The other is to use HQET effective theory with $1/m_Q$
corrections. Since both approaches have their own advantages and 
disadvantages, it would be important to have both results and 
study whether they give consistent results within quoted errors. 
I will present some of the major calculations from these two
approaches and discuss the consistency of the results.

The second problem already gives a source of errors in the
quenched calculations but it would become even more serious 
in unquenched QCD. Unfortunately, there are still no results in the 
unquenched QCD. Since the chiral limit of the form factors is 
ill-defined in the quenched QCD, all we can do with the present lattice
data is to discuss the light quark mass dependence in the intermediate 
mass regime in the quenched QCD. However, one may be able to 
give an estimate of the low energy coefficients of the chiral 
perturbation theory based on the present lattice data. 
I will briefly review new studies of quenched chiral perturbation 
theory (QChPT)  and partially quenched chiral perturbation theory 
(PQChPT)~\cite{ChPT} for $\Btopi$ form factors, which  
give a phenomenological estimate on the quenching errors and 
chiral extrapolations. These studies will be even more useful 
once new calculations in unquenched QCD will be made. 

The third problem arises because in semileptonic decays   
large energies are released to the final states 
so that the lepton pair invariant mass $q^2$ can range
from $0$ to $(m_B-m_{\pi})^2$.
However, due to the discretization errors of $O(a E)$ 
as well as the statistical errors which grow as $\sim 
\exp( {\rm const} \times (E-m_{\pi}) t )$ where $E$ is 
the energy of the pion,
lattice QCD can cover only large $q^2$ region. 
The dispersion relation is a possible solution 
to give bounds for smaller $q^2$ region \cite{Dispersive}. 
The light-cone QCD sum rule (LCSR) predicts form
factors for small $q^2$, which is complementary to the 
lattice results. I will give a comparison of  the recent LCSR results with 
the lattice results to see whether they will give consistent results. 

I also review the form factors in other processes. 
Some of the recent work on the lattice QCD calculations of the
$\Btorho$ form factors in relativistic formalism are presented.  
Very precise calculations of semileptonic form factors for $B\rightarrow D^{(*)} l\nu$  at zero recoil and the calculations of the slope of the Isgur Wise
function are presented.

\section{$\Btopi$}
The exclusive semileptonic decay $\Btopi$ determines 
the CKM matrix element $|V_{ub}|$ through the 
following formula,
\begin{eqnarray}
\frac{d\Gamma}{dq^2} &=& 
\frac{G_F^2}{24\pi^3}|\boldvec{k}_{\pi}|^2 
|V_{ub}|^2 |f^+(q^2)|^2,
\end{eqnarray}
where the form factor $f^+$ is defined as
\begin{eqnarray}
\langle\pi(k_{\pi})|\bar{q}\gamma^{\mu}b|B(p_B)\rangle
&= &
f^+(q^2)  \left[ 
    (p_B+k_{\pi})^{\mu} 
    - \frac{m_B^2-m_{\pi}^2}{q^2} q^{\mu}
  \right] 
\nonumber\\
& + &
f^0(q^2) \frac{m_B^2-m_{\pi}^2}{q^2} q^{\mu},\
\end{eqnarray}
with $q = p_B-k_{\pi}$ and $q^2 = m_B^2 + m_{\pi}^2 - 2 m_B v\cdot k_{\pi}$.
The following parameterization proposed by Burdman {\it et
al.}~\cite{Burdman} 
\begin{equation}
  \label{eq:f1f2}
  \langle\pi(k_{\pi})|\bar{q}\gamma^{\mu}b|B(v)\rangle
  = 2 \left[
    f_1(v\cdot k_{\pi}) v^{\mu} +
    f_2(v\cdot k_{\pi}) \frac{k_{\pi}^{\mu}}{v\cdot k_{\pi}}
  \right], \nonumber
\end{equation}
is also useful for discussing the heavy quark symmetry and the chiral
symmetry of the form factors in a transparent way.
\subsection{Lattice results}
Lattice calculation is possible only in limited situations.
Spatial momenta must be much smaller than the cutoff, i.e.
$|\vec{p}_B|,|\vec{k}_{\pi}| < $ 1 GeV.
This means $v \cdot k_{\pi} \equiv E_{\pi} < $ 1 GeV
or equivalently $q^2 > $ 18 GeV$^2$.
Another limitation is that due to the slowing down, 
simulations with very small light quark masses are difficult
so that usual mass range for the
light quark masses in practical simulations is $m_s/3 \leq m_q \leq m_s $ or 
$m_{\pi} = 0.4 \sim 0.8 $ GeV. Therefore in order to 
obtain physical results chiral extrapolations in the light 
quark masses are necessary.

So far all the lattice calculations 
of the form factors are done only in quenched approximation.
APE collaboration \cite{Bpi_APE} and UKQCD collaboration
computed $\Btopi$ form factors for a fine lattice with 
the inverse lattice spacing $a^{-1}\sim 2.7$ GeV.
They used relativistic formalism for the heavy quark 
and extrapolated the results of heavy-light meson around 
charm quark masses to the bottom quark mass.
Fermilab collaboration \cite{Bpi_FNAL} used the Fermilab formalism 
for the heavy quark and computed the form factors on three lattices 
with $a^{-1} = 1.2 \sim 2.6$ GeV .
JLQCD collaboration \cite{Bpi_JLQCD} computed 
the form factors using NRQCD formalism for the heavy quark 
on a $a^{-1}=1.64$ GeV. 
NRQCD collaboration  \cite{Bpi_NRQCD} also used NRQCD formalism 
for the heavy quark and an improved light quark action (D234 action)
on a anisotropic lattice with $a^{-1}=1.2$ GeV (spatial), $3.3$ GeV 
(temporal). In all of these calculations the light pseudoscalar meson 
masses are $0.4 \sim 0.8$ GeV.
\begin{figure}[here]
\begin{center}
\includegraphics[width=8cm,clip=true]{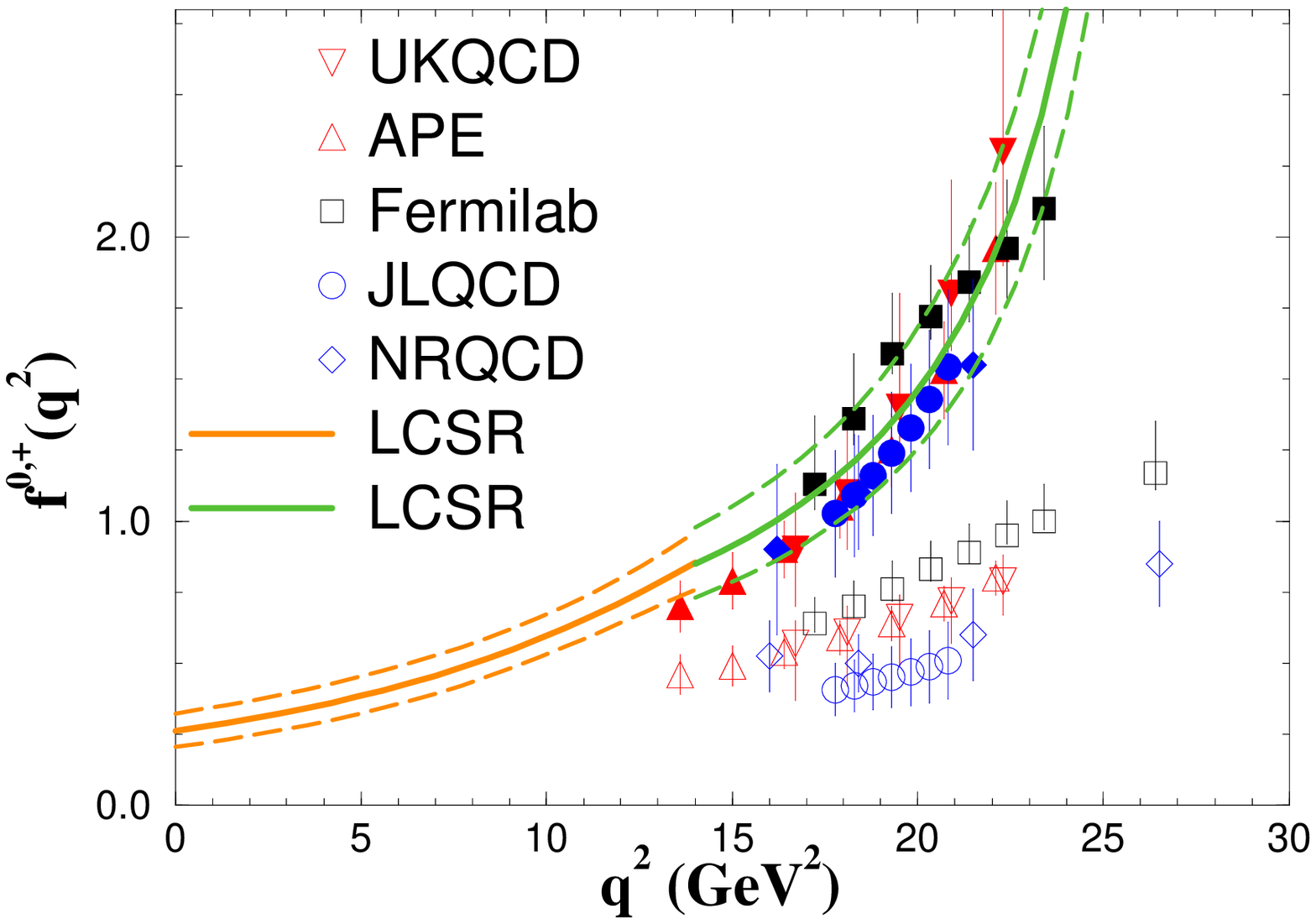}
\end{center}
\label{fig:fpf0_2}
\caption{$\Btopi$ form factors by different lattice groups.}
\end{figure}
Fig.~1 shows the result by different lattice groups,  
$f^+(q^2)$ agrees within systematic errors while 
$f^0(q^2)$ shows deviations among different methods.

The reason for the discrepancies in $f^0$ can be attributed 
to the systematic error in the chiral extrapolation
and heavy quark mass extrapolation (interpolation) error. 
In the following, we examine these errors in more detail.
Light quark mass $m_q$ dependence of form factors with fixed spatial 
momenta $ap=\frac{2\pi}{16}(1,0,0)$ is shown in 
Fig.~2.
In contrast to the JLQCD data, Fermilab data shows a significant 
increase towards the chiral limit.
 \begin{figure}[here]
\begin{center}
\includegraphics[width=8cm,clip=true]
{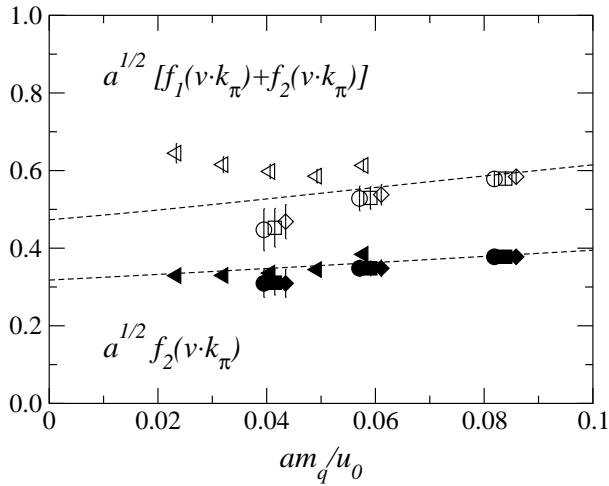}
\end{center}
\label{fig:mq_dep}
\caption{The left triangle symbols denote the Fermilab results, 
while the circle, square, and diamond symbols denote the JLQCD results
for different heavy quark masses. A clear rise towards the 
chiral limit is observed for Fermilab data.}
\end{figure}
Large difference in Fermilab results and JLQCD results for $f^0$  
in the chiral limit arises from different $m_q$ dependence, 
but the raw data for similar quark masses are not so different.
Shigemitsu {\it et al.} studied the mass dependence 
of $f_1+f_2$ and find similar behavior as JLQCD.
Further studies to clarify the light quark mass dependence are required.

Fig.~3 shows $1/M_B$ dependence of form factors
$\Phi^{0,+} \equiv \sqrt{m_B} f^{0,+}$ at $v \cdot k_{\pi}$ = 0.845 GeV
for APE and JLQCD collaboration data. 
\begin{figure}[here]
\begin{center}
\includegraphics[width=8cm,clip=true]
{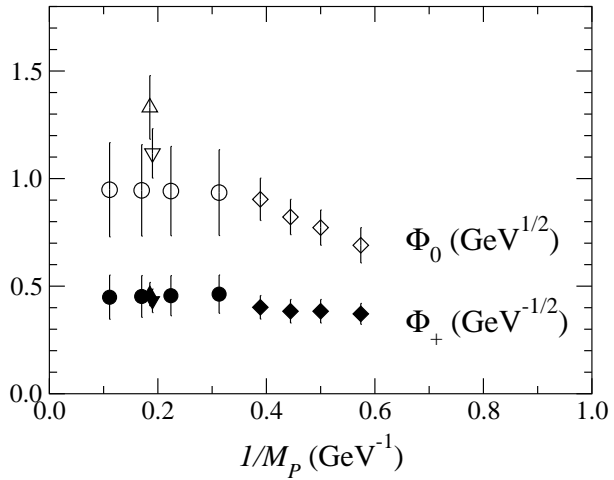}
\end{center}
\label{fig:phi_APE}
\caption{1/M scaling with fixed $v \cdot k_{\pi}$ = 0.845 GeV. 
The circle symbol denotes the JLQCD data, while the diamond symbols 
denotes the APE data.  Upward and downward triangles correspond 
the extrapolated results of APE data by linear and quadratic 
functions in $1/M$.}
\end{figure}
It is found that the difference of APE (UKQCD) vs JLQCD (NRQCD) 
for$f^0$  arises from the  extrapolation in 1/M. 
Linear extrapolation in 1/M is consistent, while the quadratic 
extrapolation gives higher value.
The quadratic extrapolation 1/M is chosen for APE's result, 
since higher value  gives better agreement with the soft pion theorem.
Simulations with static heavy quark may resolve the problem.

The error of the form factors in the present calculations is 
around 20\%. 
Some of the major errors are the  quenching error, chiral extrapolation error 
statistical error in all calculations. In addition, a large
discretization error appears in  JLQCD results and a large $1/M$
extrapolation error is contained in APE and UKQCD results.

There are several proposals to improve the form factor determination. 
The quenching error can be resolved only by performing the unquenched 
calculations. Recently, JLQCD and UKQCD collaborations has accumulated 
$n_f=2$ unquenched lattice configurations with $O(a)$-improved Wilson 
fermions and $n_f=2+1$ unquenched configurations with improved staggered 
fermions have been produced by the MILC collaboration. These unquenched QCD 
data should be applied to form factor calculations. 

In order to reduce the chiral extrapolation error, simulation with 
even smaller light quark masses are necessary. For Wilson type
fermions, simulations with $m_{\pi} < 0.4$ GeV will be very slow and 
also appearance of exceptional configuration may prevent the
simulation for very light quark mass range.
On the other hand, MILC collaboration is now carrying out simulations 
with $m_{\pi} = 0.3-0.5$  GeV, which corresponds to 
$m_q=1/5 m_s - 1/2 m_s$ \cite{ref:Mackenzie}. Since $n_f=2+1$
simulations are performed by taking the square root or quartic root of 
the Dirac operator, one possible concern might be the 
flavor symmetry breaking effect which still exists in the improved 
stagger fermions.  It would be 
important to understand how this flavor breaking affects distorts 
the chiral behavior.  Although it is possible to simulate with very 
light quark mass without encountering any theoretical problems, the 
simulation cost at this stage is very high. A breakthrough in the 
algorithm is needed.

Using the heavy quark symmetry is another way for improvements. 
Since CLEO-c experiment can measure form factors for $\Dtopi$ 
to a few percent accuracy, their results will be a good approximation
for the $\Btopi$ form factors. Then the task for lattice QCD is 
to predict the $1/m_Q$ dependence of the form factors. 
The B/D ratio $\displaystyle{\frac{d\Gamma(\Btopi)}{d\Gamma(\Dtopi)}}$ 
with the same recoil energy $\vdotk$ would be a nice quantity to measure 
on the lattice, since a large part of the statistical error, 
the perturbative error and  the chiral extrapolation errors are 
expected to cancel in this ratio  .

\subsection{QChPT and  (PQ)ChPT}

Becirevic {\it et al.} \cite{ChPT} made estimates of the quenching effect and
the chiral extrapolation error  from low energy effective meson
theory, namely QChPT and (PQ)ChPT for quenched QCD and full QCD respectively.

In their analysis the non perturbative low energy coupling constant are 
(1) $\alpha$, $g$ and $f$ for $f_B$ $B^*B\pi$ coupling and $f_{\pi}$,
(2) $L_4$ and  $L_5$ for the two $O(p^4)$ terms, and 
(3) $m_0$ and $g'$ for the two parameters of quenching effect.
The parameters are estimated by collecting all the 
knowledge from the quenched lattice QCD simulation,
the full QCD simulation, the experimental values and large N order estimation.

They found that the quenching errors 
$Q_2 \equiv (f^{full}_2-f^{quench}_2)/f^{full}_2$
and 
$Q_{12} \equiv 
((f1+f_2)^{full}-(f_1+f_2)^{quench}))/(f_1+f_2)^{full}$ 
are as large as 25\%-50\% depending on  $\vdotk$ and $g'$.

They also estimated the chiral extrapolation error 
by comparing the result with linear extrapolation and 
linear plus chiral log corrections for unquenched theory.
They found that the linear extrapolation of $m_{\pi} > 0.4 $ GeV 
data gives an overestimate of $f_{\perp} \propto f_2$ by 
2-7\% for $\vdotk=0.19$ GeV and 5-15\% for $\vdotk=0.54$ GeV.

They proposed an extrapolation strategy in the partially quenched theory 
for $m_{\pi}^{sea}>0.4$  GeV simulation data in which one first takes a 
linear extrapolation of $m_{sea}$ with fixed $m_{valence}$  and then 
make an extrapolation in $m_{valence}$ with a linear + log form. 
In this method they estimate that the chiral extrapolation error 
in unquenched theory in under 5\% level.

\subsection{$q^2$ dependence from the dispersive bound and LCSR}
Model independent bounds for the whole $q^2$ range
can be obtained with the dispersion relation, perturbative QCD,
and the lattice QCD data \cite{Dispersive}.
\begin{figure}[here]
\begin{center}
\includegraphics[width=7cm,clip=true]{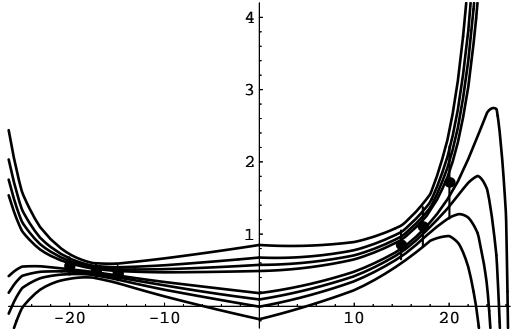}
\caption{Dispersive bound for $f^0$ and $f^+$ with UKQCD lattice data.
Model-independent QCD bounds with 90\%, 70\%, 50\% and 30\% confidence
levels are given by the pair of curves. Figure taken from 
Ref.~\cite{Dispersive}.}
\end{center}
\label{Disp}
\end{figure}
Fig.~4 shows the pioneering result by Lellouch
\cite{Dispersive}. Reducing the lattice errors or having other 
inputs would significantly improve the results. More elaborate studies 
along this line would be important. 

The light-cone QCD sum rule (LCSR)~\cite{LCSR_Ball,LCSR_Khod,LCSR_BZ}
can also give form factors for small $q^2$ region.
The basic idea is to compute the following 
matrix element $CF_v$ 
\begin{eqnarray}
CF_v &\equiv& i \int d^4 y e^{iqy} 
\langle \pi(p) | T[\bar{q}\gamma_{\mu} b](y) j_B(0) | 0 \rangle,
\end{eqnarray}
in two different methods and equate the results.
The two computational methods are :
(1) light cone expansion which is expressed by 
the pion light-cone wavefunction $\phi_{\pi}(u)$ and 
(2) the dispersion relation which takes the following 
sum of the physical poles 
\begin{eqnarray}
CF_v &\sim& \frac{m_B^2 f_B}{m_b} f^+(q^2) \frac{1}{m_B^2-p_B^2}
+ \mbox{higher poles},
\end{eqnarray}
where higher poles are suppressed by Borel transformation with M
and approximated by the light-cone expansion results 
above a threshold $s_0^2$.

The theoretical input parameters are 
the parameters $a_i$'s of light-cone wavefunctions in 
the Gegenbauer polynomial expansion,
\begin{eqnarray}
\phi_{\pi} &=& 6u(1-u) [ 1 + a_2 C_2^{3/2}(2u-1) + \cdots ],
\end{eqnarray}
the B meson decay constant $f_B$, the b quark mass $m_b$,  
the threshold $s_0^2$, and the parameter $M$ for the Borel transformation.

We here give the new results of Ref. \cite{LCSR_BZ} as an example.
It is found that the radiative correction is  about 10\% and the 
correction from higher twists ( twist 3) is $\sim$ 30\%.
The results for $q^2<14$ GeV$^2$ are well fitted by 
\begin{eqnarray}
f^+(q^2)&=&\frac{F(0)}{1-a q^2/m_B^2 + b (q^2/m_B^2)^2}.
\end{eqnarray}
Light-cone QCD sum rule results for $q^2<14$ GeV$^2$ 
can also be fitted by the pole dominance ansatz.
\begin{eqnarray}
f^+(q^2) &=& \frac{c}{1-q^2/m_{B^*}^2}
\end{eqnarray}
where $c\equiv f_{B^*} g_{BB^*\pi}/(2 m_{B^*}) = 0.414
^{+0.016}_{-0.018}$ plus systematic errors.

Fig.~\ref{fig:fpf0_2} shows the result by the light-cone QCD sum rule.
It is remarkable that the light-cone QCD sum rule give 
consistent results with lattice QCD.

\section{$B\rightarrow\rho l \nu$}
Recently, UKQCD collaboration \cite{Brho_UKQCD} and 
SPQcdR collaboration \cite{Brho_SPQcdR}
started studies of $\Btorho$ form factors.
Both collaborations use $O(a)$-improved Wilson action for the 
heavy quark and extrapolate the numerical results of $m_Q \sim m_c$ 
towards the physical b quark mass. The lattice spacings are
$a^{-1}$= 2.0 and 2.7 GeV for UKQCD and $a^{-1}$ = 2.7 and 3.7 GeV for SPQcdR.

UKQCD fits the lattice data for $q^2>$ 14 GeV$^2$
to the following form 
\begin{eqnarray}
\frac{1}{|V_{ub}|^2}\frac{d\Gamma}{dq^2}
&=&
 \frac{G_F^2 q^2 [\lambda(q^2)]^{1/2}}{192\pi^3 m_B^3} 
(a+ b (q^2-q^2_{max})), \nonumber.
\end{eqnarray}
The fit coefficients are $a = 38 ^{+8}_{-5} \pm 5$ GeV$^2$ 
and $b = 0 \pm 2 \pm 1$, where the first error is statistical 
and the second is the extrapolation error for both $a$ and $b$ .

SPQcdR collaboration obtains form factors for $q^2>$ 10 GeV$^2$.
They find the results which is consistent with the light-cone QCD 
sum rule results.
\begin{figure}[here]
\begin{center}
\includegraphics[width=8cm,clip=true]{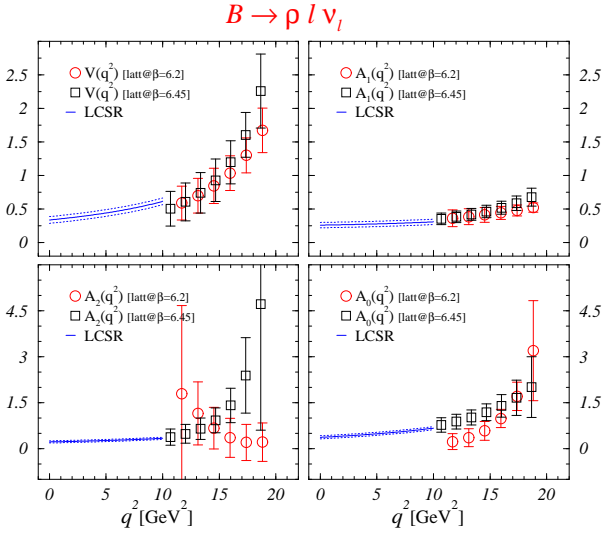}
\end{center}
\label{fig:plot_brho}
\caption{A comparison of $\Btorho$ form factors from lattices for 
$q^2 > 10$ GeV$^2$ with the light-cone sum rule (LCSR) results 
 $q^2 < 10$ GeV$^2$ region. Figure taken from Ref.~\cite{Brho_SPQcdR}.}
\end{figure}

\section{$\BtoD$}
One can extract $|V_{cb}|$ from the $\BtoD$ semileptonic decay near 
zero recoil as
\begin{eqnarray}
\frac{d\Gamma}{d\omega}(B\rightarrow D^{(*)})
&\propto &
|V_{cb}|^2 |{\cal F}_{B\rightarrow D^{(*)}}(\omega)|^2 ,
\end{eqnarray}
where $\omega\equiv v\cdot v'$ and 
${\cal F}_{\BtoD}$ are the linear combinations of form factors 
$h_{\pm},h_{A_{1,2,3}}$.
One important outcome from the heavy quark symmetry is that  
the form factor ${\cal F}_{\BtoD}$ is equal to unity at zero recoil up to 
perturbatively calculable factors and $1/m_{b,c}$ corrections 
only starts from the second order as,
\begin{eqnarray}
{\cal F}_{B\rightarrow D^{*}}(1)
&=& \eta_A \left[ 1- \frac{l_V}{(2m_c)^2}
       + \frac{2 l_A}{2m_c 2m_b} - \frac{l_P}{(2m_b)^2} \right] \nonumber\\
{\cal F}_{B\rightarrow D}(1)
&=& \eta_V \left[ 1- l_P (\frac{1}{2m_c}-\frac{1}{2m_b})^2 \right].
\end{eqnarray}
Hashimoto {\it et al.} \cite{BD_FNAL}
computed the double ratio in lattice QCD
to obtain the coefficients $l_V, l_A, l_P$
\begin{small}
\begin{equation}
\frac{\langle D|\bar{c}\gamma^0 b|\bar{B}\rangle
      \langle\bar{B}|\bar{b}\gamma^0 c|D\rangle}  
     {\langle D|\bar{c}\gamma^0 c|D\rangle
      \langle\bar{B}|\bar{b}\gamma^0 b|\bar{B}\rangle}
=
 | \left[ 1-  l_P(\frac{1}{2m_c}-\frac{1}{2m_b})^2 \right]|^2 
\end{equation}
\begin{equation}
\frac{\langle D^*|\bar{c}\gamma^0 b|\bar{B}^*\rangle
      \langle\bar{B}^*|\bar{b}\gamma^0 c|D^*\rangle}  
     {\langle D^*|\bar{c}\gamma^0 c|D^*\rangle
      \langle\bar{B}^*|\bar{b}\gamma^0 b|\bar{B}^*\rangle}  
=
 | \left[ 1-  l_V(\frac{1}{2m_c}-\frac{1}{2m_b})^2 \right]|^2 
\end{equation}
\begin{equation}
\frac{\langle D*|\bar{c}\gamma^i\gamma^5 b|\bar{B}\rangle
      \langle\bar{B}^*|\bar{b}\gamma^i\gamma^5 c|D\rangle}  
     {\langle D^*|\bar{c}\gamma^i\gamma^5 c|D\rangle
      \langle\bar{B}^*|\bar{b}\gamma^i\gamma^5 b|\bar{B}\rangle}
=
 | \left[ 1-  l_A(\frac{1}{2m_c}-\frac{1}{2m_b})^2\right]|^2 
\end{equation}
\end{small}
The deviation from unity in the double ratio 
only start from $1/m^2$ including systematic errors.
The $B\to D^*$ result is 
\begin{eqnarray}
\mathcal{F}_{B\to D^*}(1) 
&=& 0.913 {}^{+0.024}_{-0.017}\pm 0.016 {}^{+0.003}_{-0.014} 
{}^{+0.000}_{-0.016} {}^{+0.006}_{-0.014}\nonumber,
\end{eqnarray}
where the first error is statistical and the second, the third, the
fourth and the fifth  are the errors from the perturbation, the
discretization, the chiral extrapolation, and quenching.

The $B\to D$ result is 
\begin{eqnarray}
 \mathcal{F}_{B\to D}(1) 
&= & 1.058 \pm 0.016 \pm 0.003 {}^{+0.014}_{-0.005}\nonumber,
\end{eqnarray}
where the first is statistical and the second is the error from 
$m_{b,c}$ and the third is the perturbative error.

UKQCD collaboration \cite{BD_UKQCD} evaluated the $\omega$ 
dependence of the form factors $\BtoD$ from the 3 point functions
in quenched QCD. Parameterizing the form factor as 
\begin{eqnarray}
h_j(\omega) &\propto& \xi(\omega) = 1 - \rho^2 (\omega-1) + 
{\cal O}((\omega-1)^2) 
\end{eqnarray}
They obtain the slope 
\begin{eqnarray}
\rho^2 &=& 0.83 {}^{+0.15}_{-0.11} {}^{+0.24}_{-0.01},
\end{eqnarray}
where the first and the second errors are statistical 
and systematic, respectively.

UKQCD collaboration \cite{Lambda_bc_UKQCD} also computed the form factors 
$\Lambda_b \rightarrow \Lambda_c l\nu$ .
Gottlieb and Tamhankar \cite{Indiana}  also made an exploratory study 
on he same form factors and obtained $\xi(\omega)$ for finite heavy
quark mass. Both groups found that the heavy quark mass dependence is small.

\section{Summary}
$\Btopi$ form factors are computed with 20\% error
for $q^2>$ 18 GeV$^2$. 
Different approaches show good agreements in the form factors 
for the same mass range of the heavy and the light quarks, 
although the extrapolation 
in $m_q$ and $1/M_Q$ gives deviations  in $f^0$. Therefore, 
a better understanding of the quark mass dependence is required.
Studies in QChPT and PQChPT suggests that the unquenching 
error may be quite large so that lattice results in unquenched
QCD are required. Once unquenched lattice results will be 
available, studies in PQChPT suggest that chiral extrapolation 
errors can be controlled below 5\% level. 
The form factor for smaller $q^2$ range can be obtained either by the 
dispersive bounds or Light-Cone QCD sum rule with the help of 
nonperturbative inputs.  $\Btorho$ form factors are
in progress. It seems that the form factors in the range $q^2>$ 14
GeV$^2$ are feasible. $B\rightarrow D^{(*)} l \nu$ results are
established in  quenched QCD, whose extension to the unquenched calculation 
should be straightforward.

\end{document}